\documentclass[11pt]{article}
\usepackage[T2A]{fontenc}
\usepackage{ alphalph,etoolbox }
\usepackage{amsthm, amsmath, amssymb, amsbsy, amscd, amsfonts, latexsym, euscript,epsf, bm}

\usepackage{bbold, bm}
\usepackage{epic,eepic}
\usepackage{texdraw}
\usepackage[dvips]{color}
\usepackage{color}
\usepackage{ dsfont }
\usepackage{graphicx}
\usepackage{exscale,relsize}
\usepackage{authblk}
\usepackage{url}
\usepackage{enumerate}
\usepackage{graphicx}

\title{The emergence of Airy stress function in two-dimensional disordered packings of particles.}
\date{}

\author[1]{Dmitry Grinev\thanks{d.grinyev@uniyar.ac.ru}}
\affil[1]{Centre of Integrable Systems, P.G. Demidov Yaroslavl State University, Yaroslavl, Russia}

\patchcmd{\subequations}{\alph{equation}}{\alphalph{\value{equation}}}{}{}

\begin{document}

\maketitle

\begin{abstract}
The packing of hard-core particles in contact with their neighbors offers the statically determinate problem which allows 
analytical investigation of the stress tensor distribution. We construct the stress probability functional method and derive the complete set of equations for the macroscopic tensor components in two dimensions. 
For the isotropic and homogeneous two-dimensional packing the classical Euler-Cauchy and Navier equation are derived. For such packing its macroscopic stress tensor can be expressed via the Airy stress function. 
\end{abstract}

\bigskip

\section{Introduction}\label{sec1}

First-principles derivation of equations that govern transmission of stress in disordered particulate media still appears to remain a challenging problem \cite{bib1}. On one hand there are conventional theoretical approaches that are perfectly adequate for describing macroscopic mechanical behavior of soils driven by external forces \cite{bib2}. On the other hand there is experimental evidence of rheological behavior \cite{bib3}, that require statistical-mechanical treatment. There have been various attempts to develop statistical-mechanical approaches for such systems \cite{bib4}. However they have been treated with a degree of healthy skepticism by soil mechanics and geotechnical engineering communities. This is related to the fact that there are classical  problems of granular mechanics with well-known solutions. The former should be used for sanity-checks of novel theories and the latter should appear naturally in derivations. Recent attempts combined analytical methods with discrete element modeling to study Janssen effect and stress propagation in hexagonal lattice arrays of discs \cite{bib5}-\cite{bib6}. Ordered packings are convenient and popular toy-models \cite{bib7} but ultimately one must face ubiquitous disorder at various spatial scales and figure out a path leading to continuum equations.
This paper offers a first-principles approach that links classical theories for the macroscopic stress distribution in two-dimensional granular media to the interparticle equations of force and torque balance.

\section{The  constitutive relation and Newton's laws}\label{sec2}
Let us consider a static array of hard-core particles in contact with their neighbors. The packing is assumed to be an assembly of discrete rigid particles whose interactions with their neighbors are localized at point-like contacts. When a static packing of incompressible particles in contact is subjected to external forces at its boundaries, these forces are transmitted through the contact network. This network is determined by the set of contact points in our model. Therefore the description of the network of interparticle contacts is essential for the understanding of force transmission. We assume that the set of contact points $C_{i}^{\alpha \beta}$
provides the complete geometrical specification for such static packing. We define the centroid of contacts of particle $\alpha$ as
\begin{equation}
R_{i}^{\alpha}=\frac{\sum_{\beta} C_{i}^{\alpha \beta}}{z^{\alpha}}
\end{equation}
where $i = 1, . . . d$ is the Cartesian index,  $z_{\alpha}$ is the coordination number of particle $\alpha$.  The distance between particles $\alpha$ and $\beta$ is defined as the distance
between their centroids of contacts
\begin{equation}
R_{i}^{\alpha \beta}=R_{i}^{\beta}-R_{i}^{\alpha}=r_{i}^{\alpha \beta}-r_{i}^{\beta \alpha}
\end{equation}
where $r_{i}^{\alpha \beta}$ is the $i$-th component of the vector joining the centroid of contact with the contact point i.e.
\begin{equation}
\sum_{\beta} r_{i}^{\alpha \beta}=0
\end{equation}
In $d$ dimensions Newton’s laws of force and couple balance for each particle give us the system of $\frac{Nd(d+1)}{2}$ equations for the interparticle forces $f_{i}^{\alpha \beta}$
\begin{equation}
\sum_{\beta} f_{i}^{\alpha \beta}+g_{i}^{\alpha}=0
\end{equation}
\begin{equation}
f_{i}^{\alpha \beta}+f_{i}^{\beta \alpha}=0
\end{equation}
\begin{equation}
\sum_{\beta} \epsilon_{i k l} f_{k}^{\alpha \beta} r_{l}^{\alpha \beta}+c_{i}^{\alpha}=0
\end{equation}
where $g_{i}^{\alpha}$ is the external body force acting on grain $\alpha$ and $c_{i}^{\alpha}$
is the external body couple which we take to be zero. Particles are considered to be perfectly hard, perfectly rough and each particle $\alpha$ has a coordination number $z_{\alpha}= d + 1$.
The counting of the total number of equations and the number of unknowns allows us to formulate the simplest statically determinate problem.  The force moment $S_{i j}^{\alpha}$ for grain ${\alpha}$ is
\begin{equation}
S_{i j}^{\alpha}=\sum_{\beta} f_{i}^{\alpha \beta} r_{j}^{\alpha \beta}
\end{equation}
If each particle $\alpha$ has a coordination number $z_{\alpha}= d + 1$ then the macroscopic volume average of $S_{i j}^{\alpha}$ is defined by the sets of interparticle forces and contact points. 
``Coarse-graining'' a tensor in a heterogeneous medium is normally a non-trivial task but within the confines of this paper we shall use a simple expression in the form of 
\begin{equation}
\sigma_{i j}(\vec{r} )=\frac{1}{V} \sum_{\alpha}^{N} S_{i j}^{\alpha} \delta(\vec{r} - \vec{R^{\alpha}})
\end{equation}
 
This symmetric tensor has $\frac{d(d+1)}{2}$ independent components and the number of readily equations available is $d$. These are classical Euler-Cauchy equations of the macroscopic stress equilibrium which have their origin in Newton’s second law \cite{bib8}. Thus in $2D$ there is one missing equation (usually called a constitutive relation). There have been numerous attempts to discover this ``missing equation'' for static and quasi-static packings of particles \cite{bib5}-\cite{bib10}. However this paper will focus on deriving the equations, which are already well-known. The reason for this undertaking is to offer a simple yet convincing ``sanity check'' of the previously proposed method \cite{bib9} and perhaps offer a better understanding of the links between mechanical and structural characteristics at different spatial scales. The main idea of the formalism was to consider the probability functional for the set of force moments

\begin{equation}
\begin{aligned}
P\left\{S_{i j}^{\alpha}\right\}=& \mathcal{M} \int \prod_{\alpha, \beta}^{N} \delta\left(S_{i j}^{\alpha}-\sum_{\beta} f_{i}^{\alpha \beta} r_{j}^{\alpha \beta}\right) \\
& \times P\left\{\vec{f}^{\alpha \beta}\right\} \mathcal{D} \vec{f}^{\alpha \beta}
\end{aligned}
\end{equation}

where $\mathcal{M}$ is determined by the contact network. Given the fixed geometry of the contact network the probability distribution of the set of interparticle forces that satisfy the system of Newton’s equations of balance is given by
\begin{equation}
\begin{aligned}
P\left\{\vec{f}^{\alpha \beta}\right\}=& \mathcal{N} \prod_{\alpha=1, \beta=n . n .}^{N} \delta\left(\sum_{\beta} f_{i}^{\alpha \beta}+g_{i}^{\alpha}\right) \\
& \times \delta\left(\sum_{\beta} \epsilon_{i k l} f_{k}^{\alpha \beta} r_{l}^{\alpha \beta}\right) \\
& \times \delta\left(f_{i}^{\alpha \beta}+f_{i}^{\beta \alpha}\right)
\end{aligned}
\end{equation}
where $\mathcal{N}$ is given by
\begin{equation}
\begin{aligned}
\mathcal{N}^{-1}  = \int \prod_{\alpha, \beta}^{N} P\left\{\vec{f}^{\alpha \beta}\right\} \mathcal{D} \vec{f}^{\alpha \beta}
\end{aligned}
\end{equation}
The main idea is to transform $P\left\{S_{i j}^{\alpha}\right\}$ into the form
\begin{equation}
\begin{aligned}
P\left\{S_{i j}^{\alpha}\right\} = P\left\{S_{i j}^{\alpha}|force\right\}P\left\{S_{i j}^{\alpha}|geometry\right\}
\end{aligned}
\end{equation}
Let us exponentiate all delta-functions to obtain
\begin{equation}
\begin{aligned}
P\left\{S_{i j}^{\alpha}\right\} = & \mathcal{N} \prod_{\alpha=1, \beta=n . n .}^{N} e^{iA}\mathcal{D} \vec{f}^{\alpha \beta}\mathcal{D} {\zeta_{ij}}^{\alpha}\mathcal{D}\vec{\gamma}^{\alpha}\mathcal{D}\vec{\lambda}^{\alpha}\vec{\eta}^{\alpha\beta}
\end{aligned}
\end{equation}
where

\begin{equation}
\begin{aligned}
A=& \sum_\alpha \zeta_{i j}^\alpha\left(S_{i j}^\alpha-\sum_\beta f_i^{\alpha \beta} r_j^{\alpha \beta}\right)+\gamma_i^\alpha\left(\sum_\beta f_i^{\alpha \beta}-g_i^\alpha\right) \\
&+\lambda_i^\alpha\left(\sum_\beta \varepsilon_{i k l} f_k^{\alpha \beta} r_l^{\alpha \beta}\right)+\eta_i^{\alpha \beta}\left(f_i^{\alpha \beta}+f_i^{\beta \alpha}\right) .
\end{aligned}
\label{expon_A}
\end{equation}

Integrating out the fields $\vec{f}^{\alpha \beta},\vec{\lambda}^{\alpha},\vec{\eta}^{\,\alpha\beta}$ gives us

\begin{equation}
\begin{aligned}
P\left\{S_{i j}^{\alpha}\right\} = & \mathcal{N} \prod_{\alpha=1, \beta=n . n .}^{N} e^{i\left(\sum_{\alpha}^{N}S_{i j}^{\alpha}\zeta_{ij}^{\alpha}-\gamma_{i}^{\alpha}g_{i}^{\alpha}\right)}
\delta\left(\zeta_{ij}^{\alpha}r_{j}^{\alpha\beta}-\gamma_{i}^{\alpha}-\zeta_{ij}^{\beta}r_{j}^{\beta\alpha}+\gamma_{i}^{\beta}\right)
\mathcal{D} {\zeta_{ij}}^{\alpha}\mathcal{D}\vec{\gamma}^{\alpha}
\end{aligned}
\end{equation}
Let us observe that the third term in equation (\ref{expon_A}) gives $\prod_{\alpha}^{N}\delta\left(S_{ij}^{\alpha}-S_{ji}^{\alpha}\right)$ so that we can interchange labels $\alpha$ and $\beta$ in the last term of (\ref{expon_A}). Using the well-known formula for the product of delta-functions $\delta\left(x+y+z_{1}\right)\delta\left(x+y+z_{2}\right)=\delta\left(z_{1}-z_{2}\right)\delta\left(x+y+\frac{z_{1}+z_{2}}{2}\right)$ we obtain the system of linear algebraic equations for $\zeta_{ij}^{\alpha}$ and $\gamma_{i}^{\alpha}$
\begin{equation}
\begin{aligned}
\zeta_{ij}^{\alpha}r_{j}^{\alpha\beta}-\gamma_{i}^{\alpha}=\zeta_{ij}^{\beta}r_{j}^{\beta\alpha}-\gamma_{i}^{\beta}
\end{aligned}
\end{equation}
We shall use this system of equations to derive the system of discrete equations for $S_{i j}^{\alpha}$ and then use the simplest ``coarse-graining'' procedure in order to derive the system of equations for the macroscopic stress tensor in two dimensions.
\section{First coordination shell approximation for the Euler-Cauchy equation}\label{sec3}

Let us present $\zeta_{ij}^{\alpha}$ as the sum of two variables

\begin{equation}
\begin{aligned}
\zeta_{ij}^{\alpha} = \zeta_{ij}^{\alpha \,0}+\zeta_{ij}^{\alpha \,*}
\end{aligned}
\end{equation}
where $\zeta_{ij}^{\alpha \,0}$ satisfies the equation

\begin{equation}
\begin{aligned}
\zeta_{ij}^{\alpha \, 0}r_{j}^{\alpha\beta}-\gamma_{i}^{\alpha}=\zeta_{ij}^{\beta\,0}r_{j}^{\beta\alpha}-\gamma_{i}^{\beta}
\end{aligned}
\label{ECdiscrete}
\end{equation}
Thus $\zeta_{ij}^{\alpha \,*}$ satisfies the set of $\frac{zNd}{2}$ equations
\begin{equation}
\begin{aligned}
\zeta_{ij}^{\alpha \, *}r_{j}^{\alpha\beta}-\zeta_{ij}^{\beta\,*}r_{j}^{\beta\alpha} = 0
\end{aligned}
\label{Navdiscrete}
\end{equation}
This system of linear equations gives the required number $\frac{Nd(d-1)}{2}$ of constraints for the set of $S_{ij}^{\alpha}$. Let us introduce a tensor $M_{ij}^{alpha}$ as the inverse of $\sum_{\beta}R_{i}^{\alpha \beta}R_{j}^{\alpha \beta}$ and rewrite (\ref{ECdiscrete}) in the following form
 \begin{equation}
\zeta_{i j}^{\alpha \, 0}=M_{j l}^\alpha \sum_\beta R_l^{\alpha \beta}\left(\gamma_i^\alpha-\gamma_i^\beta\right)+M_{j l}^\alpha \sum_\beta R_l^{\alpha \beta} r_k^{\beta \alpha}\left(\zeta_{i k}^{\beta \, 0}-\zeta_{i k}^{\alpha \, 0}\right)
\end{equation}

We can iterate this equation further in order to link the left part to the variables corresponding to next nearest neighbours of the reference particle $\alpha$. The second
iteration propagates this equation further through the network of contacts, however if we limit this process to the first coordination shell of particle $\alpha$ then we obtain the set of $Nd$ discrete equations

\begin{equation}
\sum_\beta S_{i j}^\alpha M_{j l}^\alpha R_l^{\alpha \beta}-\sum_\beta S_{i j}^\beta M_{j l}^\beta R_l^{\beta \alpha}=g_i^\alpha
\end{equation}

The simplest ``coarse-graining'' procedure applied to this set of expressions gives us $d$ Euler-Cauchy equations for macroscopic stress tensor

\begin{equation}
\nabla_j\sigma_{ij}(\vec{r})=g_{i}(\vec{r})
\end{equation}
During this derivation we assumed that the granular  packing is homogeneous and has a constant density at the macroscopic length-scale. One can think of a more sophisticated scheme of ``coarse-graining'' method in order to explore effects of heterogeneity at  different spatial scales \cite{bib11}.

\section{The Airy stress function and Navier equation}\label{sec4}

Let us now derive the remaining $d(d-1)/2$ equations for the macroscopic stress tensor and eventually limit our analysis to the two-dimensional case. In order to obtain $P\left\{S_{i j}^{\alpha}|geometry\right\}$ let us investigate the set of equations (\ref{Navdiscrete}).  Firstly it is important to observe that (\ref{Navdiscrete}) appears to be too many equations. Because of the presence of the linear combination $\sum_{\beta} r_{i}^{\alpha \beta}=0$ there are here are many internal identities and meticulous counting shows that it indeed contain $dN$ equations. Let us sum equation (\ref{Navdiscrete}) over $\beta$ and obtain

\begin{equation}
\begin{aligned}
\sum_{\beta}\zeta_{ij}^{\beta\,*}r_{j}^{\beta\alpha} = 0
\end{aligned}
\label{Navdiscrete2}
\end{equation}

This expression can be substituted into the integral for $P\left\{S_{i j}^{\alpha}|geometry\right\}$
\begin{equation}
P\left\{S_{i j}^\alpha |geometry \right\}=\int \prod_\alpha^N \mathrm{e}^{\mathrm{i} \sum_\alpha^N S_{i j}^\alpha \zeta_{i j}^{\alpha \,*}} \delta\left(\sum_\beta \zeta_{i j}^{\beta \,*} r_j^{\beta \alpha}\right) \mathcal{D} \zeta^{\alpha\, *}
\end{equation}
The set of delta-functions containing ((\ref{Navdiscrete}) can now be exponentiated and after integration out $\zeta^{\alpha\, *}$ this gives
\begin{equation}
P\left\{S_{i j}^\alpha |geometry \right\}=\int \prod_\alpha^N \delta\left(S_{i j}^\alpha-\frac{1}{2} \sum_\beta\left(\phi_i^\beta r_j^{\alpha \beta}+\phi_j^\beta r_i^{\alpha \beta}\right)\right) \mathcal{D} \phi^\alpha
\end{equation}
where the integration over the variable $\phi^\alpha$ gives the required $Nd(d-1)/2$ constraints for $S_{i j}^\alpha$

Let us now conduct the promised sanity-check of this method in $2D$ and construct the simplest interpolation of  $\phi^{\beta}$

\begin{equation}
\phi^{\beta}_{i}=\phi^{\alpha}_{i} + R_{j}^{\alpha \beta}\nabla_{j}\phi^{\alpha}_{i}
\end{equation}
After substituting it into (\ref{Navdiscrete2}) and summing over $\beta$ we have

\begin{equation}
S_{i j}^{\alpha}=\nabla_{k} \phi_{i}^{\alpha} \sum_{\beta} R_{k}^{\alpha \beta} r_{j}^{\alpha \beta}
\end{equation}
The previously employed ``coarse-graining'' method gives the macroscopic stress tensor
\begin{equation}
\sigma_{i j}=\frac{1}{V} \sum_{\alpha}^{N} \sum_{\beta} r_{j}^{\alpha \beta} R_{k}^{\alpha \beta} \nabla_{k} \phi_{i}^{\alpha}=F_{j k} \nabla_{k} \phi_{i}
\end{equation}
where $F_{ij}=\sum_{\beta}R_{i}^{\alpha \beta} R_{j}^{\alpha \beta}$ is the so called fabric tensor.
We can now eliminate $\phi_{i}$ and obtain for the isotropic packing of particles the well-known Navier equation which imposes kinematic compatibility on the stresses \cite{bib12}
\begin{equation}
\frac{\partial^{2} \sigma_{x x}}{\partial y^{2}}+\frac{\partial^{2} \sigma_{y y}}{\partial x^{2}}-2 \frac{\partial^{2} \sigma_{x y}}{\partial x \partial y}=0
\end{equation}
This equation implies that the stress tensor components can be expressed in terms of the Airy function \cite{bib12}. In 3-D there are mathematical challenges with the ``coarse-graining'' procedure which yet to be resolved.
\section{Conclusion}\label{sec4}
The application of our method allows the derivation of the classical Euler-Cauchy and Navier equations for an isotropic and homogeneous two-dimensional packing. This appears to be a much-needed sanity-check of the first-principles approach. Further development of this framework is in progress and have a potential to be applicable to quasi-static and three-dimensional problems of stress transmission in particulate media. 
\section{Conflict of Interest} 
The author declares that he has no conflict of interest.
\section{Acknowledgments}

The work on sections $3$ and $4$ was supported by the Russian Science Foundation (grant No. $21-71- 30011$). The work on sections $1$ and $2$ was carried out within the framework of a development programme for the Regional Scientific and Educational Mathematical Center of the Yaroslavl State University with financial support from the Ministry of Science and Higher Education of the Russian Federation (Agreement on provision of subsidy from the federal budget No. $075-02-2022-886$).

\end{document}